\documentclass[showpacs,preprintnumbers,amsmath,amssymb,nofootinbib]{revtex4}
\usepackage{color}
\usepackage{epsf}
\usepackage{amssymb}
\usepackage{amsfonts}
\usepackage{latexsym}
\usepackage{mathrsfs}
\usepackage{graphicx,epsf}

\def\nn{\nonumber}
\def\be{\begin{equation}}
\def\ee{\end{equation}}
\def\bea{\begin{eqnarray}}
\def\eea{\end{eqnarray}}
\def\la{\langle}
\def\ra{\rangle}
\def\bx{{\bf x}}

\def\bk{{\bf k}}
\def\bkp{{\bf k'}}

\def\fNL{{f_{NL}}}

\def\bkone{\mathbf k_1}
\def\bktwo{\mathbf k_2}
\def\bkthree{\mathbf k_3}
\def\bkfour{\mathbf k_4}
\def\bkt{\mathbf k_t}
\def\picube{(2\pi)^3}
\def\Mp{m_p}
\def\F{\mathcal{F}}

\newcommand{\sdelta}[1]{\!\delta^{\,3}(\mathbf{#1})}

\begin{document}
\title{Primordial trispectrum from inflation}
\author{Christian T.~Byrnes$^1$, Misao Sasaki$^2$ and David Wands$^1$}

\affiliation{$^1$Institute of Cosmology and Gravitation, Mercantile House, University of
Portsmouth, Portsmouth~PO1~2EG, United Kingdom \\}

\affiliation{$^2$Yukawa Institute for Theoretical Physics, Kyoto University, Kyoto
606-8503, Japan\\}

\date{\today}

\pacs{98.80.Cq \hfill PU-ICG-06, astro-ph/0611075}

\begin{abstract}

We use the $\delta N$-formalism to describe the leading order
contributions to the primordial power spectrum, bispectrum and
trispectrum in multiple-field models of inflation at leading order in
a perturbative expansion.
In slow-roll models where the initial field fluctuations at
Hubble-exit are nearly Gaussian, any detectable non-Gaussianity is
expected to come from super-Hubble evolution. We show that the
contribution to the primordial trispectrum can be described by two
non-linearity parameters, $\tau_{NL}$ and $g_{NL}$, which are
dependent upon the second and third derivatives of the local expansion
with respect to the field values during inflation.

\end{abstract}

\maketitle

\section{Introduction}

The simplest models of inflation predict a quasi scale invariant spectrum of nearly
Gaussian, adiabatic perturbations \cite{liddleandlyth}. While observations are consistent
with this picture, it is important to measure any deviations from these predictions, in
order to constrain the many models of inflation. Non-Gaussianity is a potentially
powerful discriminant between different models. Currently most efforts to constrain the
non-Gaussianity have focused on the 3-point function of perturbations, the bispectrum
\cite{Verde:1999ij,Wang:1999vf,Komatsu:2001rj}. However the 4-point function, or
trispectrum can also be constrained by increasingly accurate measurements
\cite{Okamoto:2002ik,Bartolo:2005fp,Kogo:2006kh}.

The $\delta N$-formalism \cite{Starobinsky:1986fx,Sasaki:1995aw,Lyth:2004gb,Lyth:2005fi}
identifies the primordial curvature perturbation with a perturbation in the local value
of the integrated expansion,
\begin{equation}
 \label{localN}
N(x) = \int_{t_i}^{t_p} H(x,t) dt \,,
\end{equation}
with respect to the expansion in the background spacetime
\begin{equation}
\bar{N} = \int_{t_i}^{t_p} \bar{H}(t) dt \,.
\end{equation}
This enables one to calculate the non-linear primordial curvature perturbation (e.g., at
the epoch of primordial nucleosynthesis) in terms of initial scalar field fluctuations
during inflation \cite{Lyth:2005fi}. In particular the primordial curvature perturbation
on uniform-density hypersurfaces, $\zeta$, corresponds to the perturbation in the local
expansion defined with respect to an initial spatially flat hypersurface
\cite{Lyth:2004gb}. In the long-wavelength limit where spatial gradients and anisotropic
shear becomes small, the integrated expansion along a worldline can be calculated from
solutions to the unperturbed Friedmann equation.

We have
\begin{equation}
\zeta = N(\phi^A) - \bar{N} \,,
\end{equation}
where $\phi^A$ represents the initial values for the scalar fields on the
initial spatially-flat hypersurface. If we decompose the local scalar
field values into a homogeneous background value plus perturbation,
\begin{equation}
\phi_A = \bar\phi_A + \varphi_A \,,
\end{equation}
we can write the curvature perturbation as a Taylor expansion
\begin{equation}
\label{Taylor} \zeta = \sum_{n} \frac{1}{n!} N_{A_1A_2\ldots A_n} \varphi^{A_1}
\varphi^{A_2} \ldots \varphi^{A_n} \,.
\end{equation}
There is an implicit sum over the fields for all of the repeated indices $A_i$. We are
thus able to describe the non-Gaussianity of the primordial curvature perturbation given
an initial distribution of scalar field perturbations. In practice, scalar fields during
slow-roll inflation are expected to have a nearly Gaussian distribution during inflation,
at least on scales slightly larger than the Hubble scale, i.e., shortly after Hubble exit
\cite{Seery:2005gb}. In single-field inflation this is sufficient to guarantee that the
non-Gaussianity of the primordial curvature perturbation is very small
\cite{Maldacena:2002vr,Bartolo:2004if,Seery:2005wm}. It is suppressed by slow-roll
parameters since one can express derivatives of the expansion in terms of slow-roll
parameters. However in multi-field models the perturbed expansion cannot necessarily be
related to slow-roll parameters at Hubble-exit, and the primordial curvature
perturbations may deviate strongly from a Gaussian distribution.

%In this paper we explore how the distribution of the primordial
%curvature perturbation may be expressed in terms of the distribution
%of the initial field perturbations. We start by giving the
%leading-order terms for the n-point correlation functions from
%multi-field inflation in real space, and the power-spectrum,
%bispectrum and, for the first time, the trispectrum in Fourier
%space.

%We proceed to include higher-order corrections to the n-point correlation function in
%real space, showing that we can sum over all higher-order terms in the Taylor expansion
%to derive a simpler expression in terms of the average values of the derivatives of the
%expansion with respect to the fields.

\section{Perturbative expansion}

In linear perturbation theory the initial free-field fluctuations describe Gaussian
random fields, which we will denote by $\varphi^A_1(t,x^i)$. In a perturbative expansion
higher-order interactions and non-linear evolution lead to higher-order terms which are
non-Gaussian
\begin{equation}
\varphi^A = \varphi^A_1 + \frac12 \varphi^A_2 + \frac16 \varphi^A_3 +\cdots\,.
\end{equation}

%We will assume that the scalar field distributions can be decomposed
%into homogeneous background values, a first-order perturbation with a
%Gaussian distribution, and higher-order terms that may be non-Gaussian
%and are thus expected to be small.  Specifically we split the field
%into homogeneous part plus perturbation
%%
%\bea \phi=\phi_0(t)+\varphi(t,x^i)\,, \eea
%%
%where the perturbation is decomposed into gaussian, ($\varphi_g$), and
%non-gaussian parts
%%
%\bea \varphi=\varphi_g+\frac12\varphi_b+\frac16\varphi_t+\cdots \eea
%%
{}From the definition of Gaussian statistics, the $n$-point correlator
of the Gaussian fluctuation is zero for odd $n$ and can be reduced to
(disconnected) products of two-point functions for even $n$.

\subsection{Power spectrum}

%The 2-point function at leading-order is given by
%\begin{equation}
%\la \varphi^A(\bx) \varphi^B(\by) \ra = \delta^{AB} G(\bx-\by) \,,
%\end{equation}
%where $G$ is second-order and we have assumed that the distribution is
%spatially homogeneous. We will write the 3-point function as
%\begin{equation}
%\la \varphi^A(\bx) \varphi^B(\by) \varphi^C(\bz) \ra = \ldots
%\end{equation}

%The Fourier transform of the fields is given by
%\begin{equation}
%\varphi_\bk = \int d^3k \varphi e^{-i\bk.\bx} \,.
%\end{equation}

In Fourier space the power spectrum is given by
\begin{equation}
\la \varphi_\bk^A \varphi_\bkp^B \ra = C^{AB}(k)\picube \sdelta{\bk+\bkp} \,.
\end{equation}
This is second and higher order in a perturbative expansion.

At leading order in the field perturbations and in the slow-roll limit the fluctuations
are independent and we have
\begin{equation}\label{Cfreefields}
C^{AB}(k) = \delta^{AB} P(k) \,,
\end{equation}
where $\delta^{AB}$ is the Kronecker delta-function,
and the variance per logarithmic interval in $k$-space is given by
\begin{equation}
{\cal P}(k) = \frac{4\pi k^3}{(2\pi)^3} P(k)=\left(\frac{H_*}{2\pi}\right)^2\,,
\end{equation}
where the Hubble parameter $H$ is evaluated at Hubble-exit, $k=(aH)_*$. At zeroth order
in slow-roll parameters, $\mathcal{P}$ is independent of wavenumber, i.e.~we have a scale
invariant spectrum for the field fluctuations.

In general the fields are correlated at Hubble-exit, at first order in the slow-roll
parameters. In the case of two field inflation, using the methods of \cite{byrnes}, (see
also \cite{vanTent:2003mn}), we find for the power spectra and cross-correlation
respectively,
\bea  C^{11}&=&P_{\varphi_1}=\frac{\picube}{4\pi k^3} \left(\frac{H_*}{2\pi}\right)^2
\left[1-2\frac{\dot{H}}{H^2}+
C\left(4\frac{\dot{H}}{H^2}+\frac{1}{\Mp^2}\frac{\dot{\phi}_1^2}{H^2}-
\frac{1}{\Mp^2}\frac{\dot{\phi}_2^2}{H^2}-2\Mp^2\frac{V_{\phi_1\phi_1}}{V}\right)\right]\,,
\\ C^{22}&=&P_{\varphi_2}=\frac{\picube}{4\pi k^3} \left(\frac{H_*}{2\pi}\right)^2
\left[1-2\frac{\dot{H}}{H^2}+
C\left(4\frac{\dot{H}}{H^2}+\frac{1}{\Mp^2}\frac{\dot{\phi}_2^2}{H^2}-
\frac{1}{\Mp^2}\frac{\dot{\phi}_1^2}{H^2}-2\Mp^2\frac{V_{\phi_2\phi_2}}{V}\right)\right]\,,
\\ C^{12}&=&\frac{\picube}{4\pi k^3} \left(\frac{H_*}{2\pi}\right)^2C
\left(\frac{2}{\Mp^2}\frac{\dot{\phi}_1\dot{\phi}_2}{H^2}-2\Mp^2\frac{V_{\phi_1\phi_2}}{V}\right)\,,
\eea
where $C=2-\ln2-\gamma\simeq 0.7296$ and $\gamma$ is the Euler-Mascheroni constant.
Overdots represent derivatives with respect to time, and $V_{\phi_1}=\partial V/\partial
\phi_1 $.

\subsection{Bispectrum}

The first signal of non-Gaussianity comes from the bispectrum, which at lowest order is
\bea \langle\varphi^A\varphi^B\varphi^C\rangle = \langle\varphi_1^A
\varphi_1^B\varphi_2^C\rangle+\mathrm{perms}\,, \eea
and therefore is fourth order in perturbations.

The bispectrum of the distribution is given by
\begin{equation}
\la \varphi^A_{{\mathbf k_1}}\,\varphi^B_{{\mathbf k_2}}\, \varphi^C_{{\mathbf k_3}}  \ra
\equiv B^{ABC}(k_1,k_2,k_3) \picube \sdelta{\bkone+\bktwo+\bkthree}\,.
\label{defbispectrum}
\end{equation}
This is fourth and higher order. This was originally calculated by Maldacena,
\cite{Maldacena:2002vr} for single field inflation, and by Seery and Lidsey,
\cite{Seery:2005gb}, for multiple fields. They show that it only depends on the amplitude
of the $k$ vectors. Specifically they calculate the quantity $\mathcal{A}^{ABC}$ at
leading order in slow roll, which is related to $B^{ABC}$ by
\bea B^{ABC}(k_1,k_2,k_3) = \frac{4\pi^4}{\prod
k_i^3}\mathcal{P}^2\mathcal{A}^{ABC}(k_1,k_2,k_3)\,. \eea
They find  $\mathcal{A}^{ABC} \sim \mathcal O(\epsilon^{1/2})$
so it vanishes in the slow-roll limit.

\subsection{Trispectrum}\label{trispectrumfields}

{}From Wick's theorem, the first-order, Gaussian perturbations do not contribute to the
connected part of the four-point function,
\bea \langle\varphi_1^A\varphi_1^B\varphi_1^C\varphi_1^D\rangle_c=0\,. \eea
Note there is a disconnected part of the four-point function which is a product of two
two-point functions which is only fourth order in slow roll and arises even for purely
Gaussian statistics. This disconnected term is only non-zero if, e.g.,~the momenta
satisfy $\bkone+\bktwo=0$ and $\bkthree+\bkfour=0$.

The second-order field perturbations, $\varphi^A_2$, are generated from the product (or
the convolution in Fourier space) of two first order variables that have Gaussian
distributions, so they do not contribute to the (connected) four-point function at lowest
possible order,
\bea  \langle\varphi_1^A\varphi_1^B\varphi_1^C\varphi_2^D\rangle_c=0\,. \eea
Hence the leading order contribution to the connected four-point function comes from two
terms, \cite{Okamoto:2002ik},
\bea \langle\varphi_1\varphi_1\varphi_1\varphi_3\rangle_c \qquad \rm{and} \qquad
\langle\varphi_1\varphi_1\varphi_2\varphi_2\rangle_c\,. \eea
Both of these terms are sixth-order in perturbations.

%[MAYBE DONT NEED THIS PARAGRAPH] In the expansion of the trispectrum we need to go as far
%as the 6-point function. The leading order part of the 5-point function comes from the
%the disconnected part
%%
%\bea \langle\varphi_g\varphi_g\rangle\langle\varphi_g\varphi_g\varphi_b\rangle \eea
%%
%and is 6'th order in slow roll. The leading order part of the 6th point function is from
%the disconnected, purely Gaussian part,
%%
%\bea
%\langle\varphi_g\varphi_g\rangle\langle\varphi_g\varphi_g\rangle\langle\varphi_g\varphi_g\rangle\,,
%\eea
%%
%and again is sixth order. We do not need to consider the seventh- or higher-point
%functions since these are eighth or higher order.

In Fourier space the connected part of the four-point function is sixth order, and given
by
\bea \la \varphi^A_{{\mathbf k_1}}\,\varphi^B_{{\mathbf k_2}}\, \varphi^C_{{\mathbf k_3}}
\varphi^D_{{\mathbf k_4}} \ra_c &\equiv& T^{ABCD}(\bkone,\bktwo,\bkthree,\bkfour) \picube
\sdelta{\bkone+\bktwo+\bkthree+\bkfour}\,. \label{deftrispectrum} \eea
This was recently calculated at leading order by Seery, Lidsey and Sloth
\cite{Seery:2006vu}. They show this quantity depends on both the magnitude and direction
of the $\bk$ vectors, specifically it depends on $k_i=|\mathbf{k}_i|$ and
$\mathbf{k}_i\cdot\mathbf{k}_j$.

\section{The primordial n-point functions}

So far we have calculated the two-, three- and four-point function of the field
fluctuations. To link these to observations, we need to calculate the n-point functions
of the primordial curvature perturbation $\zeta$. We do this using the $\delta N$
expansion for $\zeta$, (\ref{Taylor}).

\subsection{The primordial power spectrum}

At leading order the primordial power spectra depends purely on $\zeta_1$, from
Eq.~(\ref{Taylor}),
\bea\label{zeta1} \zeta_1=N_A\varphi_1^A\,. \eea
The power spectrum is thus
\begin{equation}
\la \zeta_\bk \zeta_\bkp \ra = P_\zeta(k)\picube \sdelta{\bk+\bkp} \,,
\end{equation}
where
\begin{equation}\label{PzetaC}
P_\zeta(k) = N_A N_B C^{AB}(k) \,.
\end{equation}
In the slow-roll limit, (\ref{Cfreefields}), this reduces to
\bea\label{PzetaandP} P_\zeta(k)=N_AN^AP(k)\,. \eea

\subsection{The primordial bispectrum}

To leading order in the field perturbations, the 3-point function of the curvature
perturbations depends on $\zeta_1$, (\ref{zeta1}) and
\bea\label{zeta2}  \zeta_2=N_A\varphi_2^A+ N_{AB}\varphi_1^A\varphi_1^B\,. \eea
The primordial bispectrum is thus
\bea\label{3zeta} \la \zeta_{\bkone}\zeta_{\bktwo}\zeta_{\bkthree}\ra =&& N_A N_B N_C
\la\varphi^A_{\bkone}\varphi^B_{\bktwo}\varphi^C_{\bkthree}\ra \nn
\\ &&  +\frac12N_{A_1A_2}N_B N_C  \left[ \la\left(\varphi^{A_1}*\varphi^{A_2}\right)_{\bkone}
\varphi^B_{\bktwo}\varphi^C_{\bkthree}\ra + (2\,\, \rm{perms})\right]\,, \eea
where $'*'$ denotes the convolution, defined by
\bea\label{convolutiondefn} \left(\varphi^A*\varphi^B\right)_{\mathbf{k}} =
\frac{1}{\picube}\int d^3k'\varphi^A_{\mathbf{k}-\mathbf{k}'}\varphi^B_{\mathbf{k}'}\,.
\eea
Hence the bispectrum of the curvature perturbation is
\begin{equation}
\langle\zeta_{{\mathbf k_1}}\,\zeta_{{\mathbf k_2}}\, \zeta_{{\mathbf k_3}}\rangle \equiv
B_\zeta( k_1,k_2,k_3) \picube \sdelta{{\mathbf k_1}+{\mathbf k_2}+{\mathbf k_3}} \,,
\end{equation}
where to leading order \cite{Allen:2005ye}
\bea B_\zeta(k_1,k_2,k_3) &=& N_A N_B N_C B^{ABC}(k_1,k_2,k_3) \nn \\
&&+ N_A N_{BC} N_D \left[ C^{AC}(k_1) C^{BD}(k_2) + C^{AC}(k_2) C^{BD}(k_3) + C^{AC}(k_3)
C^{BD}(k_1) \right] \,. \label{zetabispectrum} \eea

{}In the slow-roll limit we can write the bispectrum as \cite{Vernizzi:2006ve}
\bea B_{\zeta}(k_1,k_2,k_3)=4\pi^4\frac{\sum_i k_i^3}{\prod_i k_i^3}
\mathcal{P}^2_{\zeta} \left( \frac{-1}{4\Mp^2N_C N^C}\frac{\F(k_1,k_2,k_3)}{\sum_i k_i^3}
+ \frac{N_{AB}N^AN^B}{\left(N_C N^C\right)^2} \right)\,, \eea
where the form factor $\F$ is defined by
\bea \F(k_1,k_2,k_3)=\sum_i k_i^3-\sum_{i\neq j}k_i k_j^2 -8\frac{\sum_{i>j}k_i^2
k_j^2}{k_1+k_2+k_3}\,. \eea

\subsection{The primordial trispectrum}

{}From the discussion in sec.~\ref{trispectrumfields} the four-point function of the
curvature perturbation at leading order will depend on $\zeta_1$, (\ref{zeta1}),
$\zeta_2$, (\ref{zeta2}), and
\bea\label{zeta3} \zeta_3=N_A\varphi_3^A
+N_{AB}\left(\varphi_1^A\varphi_2^B+\varphi_2^A\varphi_1^B\right) +
N_{ABC}\varphi_1^A\varphi_1^B\varphi_1^C \eea
The four-point function at leading order is
\bea\label{4zeta} \la \zeta_{\bkone}\zeta_{\bktwo}\zeta_{\bkthree}\zeta_{\bkfour}\ra_c
&=& N_A N_B N_C N_D
\la\varphi^A_{\bkone}\varphi^B_{\bktwo}\varphi^C_{\bkthree}\varphi^D_{\bkfour}\ra_c \nn
\\ &&  +\frac12N_{A_1A_2}N_B N_C N_D \left[ \la\left(\varphi^{A_1}\ast\varphi^{A_2}\right)_{\bkone}
\varphi^B_{\bktwo}\varphi^C_{\bkthree}\varphi^D_{\bkfour}\ra + (3\,\, \rm{perms})\right] \nn \\
&& + \frac14N_{A_1A_2}N_{B_1B_2}N_C N_D
\left[\la\left(\varphi^{A_1}\ast\varphi^{A_2}\right)_{\bkone}
\left(\varphi^{B_1}\ast\varphi^{B_2}\right)_{\bktwo}\varphi^C_{\bkthree}\varphi^D_{\bkfour}\ra
+ (5\,\, \rm{perms})\right] \nn  \\ && +\frac{1}{3!}N_{A_1A_2A_3}N_B N_C N_D
\left[\la\left(\varphi^{A_1}\ast\varphi^{A_2}\ast\varphi^{A_3}\right)_{\bkone}
\varphi^B_{\bktwo}\varphi^C_{\bkthree}\varphi^D_{\bkfour}\ra + (3\,\,
\rm{perms})\right]\,. \eea
All of the terms shown are sixth order in the field perturbations.

The first term of the expansion above is the intrinsic 4-point function of the fields, as
calculated, \cite{Seery:2006vu}. The disconnected part of this term would only give a
contribution if the sum of any two $k$ vectors is zero, e.g.~if $\bkone+\bktwo=0$. We
will exclude this case, which is equivalent to neglecting parallelograms of the
wavevectors.

The second term of (\ref{4zeta}) consists of permutations of terms of the form
\bea\label{secondterm}  \frac12N_{A_1A_2}N_B N_C N_D \frac{1}{\picube} \int d^3q \la
\varphi^{A_1}_{\mathbf q } \varphi^{A_2}_{\mathbf{k}_1-\mathbf{q}}
\varphi^B_{\bktwo}\varphi^C_{\bkthree}\varphi^D_{\bkfour}\ra \eea
This five-point function is zero for the first-order, Gaussian, perturbations, hence the
leading order contribution is sixth-order, due to the second order contribution of one of
the fields. Hence we use Wick's theorem to split the 5-point function in to lower point
functions. There is no contribution to (\ref{secondterm}) from the split into a
four-point and a one-point function. Only the split into a two-point and three-point
function gives a contribution. However the first possible contraction in
(\ref{secondterm}), $\la\varphi^{A_1}_{k_1-q}\varphi^{A_2}_q\ra$, does not contribute
since it is only non-zero when $k_1=0$. Therefore we can reduce the above term into a
power spectra and a trispectrum in 6 different ways, which gives three distinct pairs of
terms. In total the second term of (\ref{4zeta}) is
\bea N_{A_1A_2}N_B N_C N_D \left[ C^{A_1B}(k_1)B^{A_2BC}(k_{12},k_3,k_4)+ (11\,\,
\rm{perms})\right]\picube\sdelta{\bkt}\,, \eea
where we use the shortened notation $k_{ij}=|\mathbf{k}_i+\mathbf{k}_j|$ and
$\bkt=\bkone+\bktwo+\bkthree+\bkfour$. The 12 permutations come from having 3 distinct
choices for the indices of the wavenumber $k_{ij}$ (only three distinct choices because
$k_{ij}=k_{ji}$ and $k_{12}=k_{34}$ etc). We then choose which two wavenumbers form the
remaining arguments of $B^{ABC}$, either $k_i$, $k_j$ or the other pair of wavenumbers,
and finally we choose which of the two indices $i$ or $j$ is attached to the wavenumber
$k_i$ that is the argument of C.

Continuing this argument for the second and third terms of (\ref{4zeta}), we find the
connected part of the trispectrum of the curvature perturbation is
\begin{equation}
\la \zeta_{{\mathbf k_1}}\,\zeta_{{\mathbf k_2}}\, \zeta_{{\mathbf k_3}} \zeta_{{\mathbf
k_4}} \ra_c \equiv T_\zeta({\mathbf k_1},{\mathbf k_2},{\mathbf k_3}, {\mathbf k_4})
\picube \sdelta{{\mathbf k_1}+{\mathbf k_2}+{\mathbf k_3} +{\mathbf k_4}}\,,
\end{equation}
where
\bea\label{trispectrumfull}
T_\zeta(\bkone,\bktwo,\bkthree,\bkfour) &=&N_AN_BN_CN_D T^{ABCD}(\bkone,\bktwo,\bkthree,\bkfour) \nn \\
&& + N_{A_1A_2}N_B N_C N_D \left[ C^{A_1B}(k_1)B^{A_2BC}(k_{12},k_3,k_4)+ (11\,\,
\rm{perms})\right] \nn
\\ && + N_{A_1A_2}N_{B_1B_2}N_CN_D \left[C^{A_2B_2}(k_{13})C^{A_1C}(k_3)C^{B_1D}(k_4) +(11\,\,
\rm{perms})\right]  \nn \\ && +N_{A_1A_2A_3}N_BN_CN_D
\left[C^{A_1B}(k_2)C^{A_2C}(k_3)C^{A_3D}(k_4)+(3\,\,\rm{perms})\right]\,. \eea

\section{Gaussian scalar fields}

If the scalar field perturbations are independent, Gaussian random fields, as we expect
shortly after Hubble-exit during inflation in the slow-roll limit
\cite{Maldacena:2002vr,Seery:2005gb} then the bispectrum for the fields, $B^{ABC}$, and
connected part of the trispectrum, $T^{ABCD}$, both vanish.

In this case the bispectrum of the primordial curvature perturbation
(\ref{zetabispectrum}) at leading (fourth) order, can be written as
\bea\label{fNLmultifielddefn} B_\zeta(\bkone,\bktwo,\bkthree) &=&
 \frac65 \fNL \left[ P_\zeta(k_1) P_\zeta(k_2) + P_\zeta(k_2)
P_\zeta(k_3) + P_\zeta(k_3) P_\zeta(k_1) \right] \,. \eea
where the dimensionless non-linearity parameter\footnote{Some papers
  use a different sign convention for
  $f_{NL}$. For example, Refs.~\cite{Lyth:2005fi,Vernizzi:2006ve} use
  the opposite sign convention.} is given by \cite{Lyth:2005fi}
\begin{equation}
 \label{fNLmultifield}
 \fNL = \frac{5}{6} \frac{N_A N_B N^{AB}}{\left(N_C N^C\right)^2} \,.
\end{equation}

The trispectrum (\ref{trispectrumfull}) in this case reduces to
\bea\label{trispectrumgauss}
 T_\zeta (\bkone,\bktwo,\bkthree,\bkfour) &=&
N_{AB}N^{AC}N^BN_C\left[P(k_{13})P(k_3)P(k_4)+(11\,\,\rm{perms})\right] \nn \\
&&+N_{ABC}N^AN^BN^C\left[P(k_2)P(k_3)P(k_4)+(3\,\,\rm{perms})\right]\,, \eea
Hence we can write the trispectrum as
\bea\label{tauNLgNLdefn} T_\zeta (\bkone,\bktwo,\bkthree,\bkfour) &=&
\tau_{NL}\left[P_\zeta(k_{13})P_\zeta(k_3)P_\zeta(k_4)+(11\,\,\rm{perms})\right] \nn \\
&&+\frac{54}{25}g_{NL}\left[P_\zeta(k_2)P_\zeta(k_3)P_\zeta(k_4)+(3\,\,\rm{perms})\right]\,.
\eea
where comparing the above two expressions, and using (\ref{PzetaandP}) we see
\bea
 \label{tauNLmultifield}
  \tau_{NL}&=&\frac{N_{AB}N^{AC}N^BN_C}{(N_DN^D)^3}\,, \\
 \label{gNLmultifield}
  g_{NL}&=&\frac{25}{54}\frac{N_{ABC}N^AN^BN^C}{(N_DN^D)^3}\,.
 \eea
The expression for $\tau_{NL}$ from multiple fields was given in the arXiv version of
\cite{Alabidi:2005qi}. Note that we have factored out products in the trispectrum with
different $k$ dependence in order to define the two $k$ independent non-linearity
parameters $\tau_{NL}$ and $g_{NL}$. This gives the possibility that observations may be
able to distinguish between the two parameters \cite{Okamoto:2002ik}.

\subsection{Single field dependence}

In many cases there is single direction in field-space, $\varphi$, which is responsible
for perturbing the local expansion, $N(\varphi)$, and hence generating the primordial
curvature perturbation, $\zeta$. For example this would be the inflaton field in single
field models of inflation, or it could be the late-decaying scalar field in the curvaton
scenario \cite{curvaton}.

In the case where a single field dominates, the curvature perturbation
(\ref{Taylor}) is given by
\bea
 \zeta
%  &=& \sum_{n=1}^{\infty}\frac{1}{n!}\zeta_{n}\nonumber\\
  &=& N' \varphi + \frac12 N'' \varphi^2 + \frac16 N''' \varphi^3 + \cdots\,,
\eea
where we use the shorthand $N'=dN/d\varphi$.
If in addition we assume that the field perturbation is purely
Gaussian, $\varphi=\varphi_1$, then the non-Gaussianity of the
primordial perturbation has a simple ``local form''
where the full non-linear perturbation at any point in real space,
$\zeta(\bx)$, is a local function of a single Gaussian random field,
$\varphi_1$.
Thus we can write \cite{Komatsu:2001rj,Sasaki:2006kq}
\bea
 \label{fandgdefn}
 \zeta=\zeta_1+\frac35f_{NL}\zeta_1^2+\frac{9}{25}g_{NL}\zeta_1^3+\cdots\,,
 \eea
where $\zeta_1$ is Gaussian because it is directly proportional to the initial Gaussian
field perturbation, $\varphi_1$, and the dimensionless non-linearity parameters, $f_{NL}$
and $g_{NL}$, are given by
\bea
 \label{fNL1field}
  f_{NL}&=&\frac56\frac{N''}{(N')^2}\,, \\
 \label{gNL1field}
  g_{NL}&=&\frac{25}{54}\frac{N'''}{(N')^3}\,,
\eea

The numerical factors in Eq.~(\ref{fandgdefn}) arise
because the original definition is given in terms of the
%curvature perturbation $\zeta$ is related to the
Bardeen potential on large scales (in the matter dominated era, md),
$\Phi_{H\rm{md}}=(3/5)\zeta_1$, so we have
\cite{Komatsu:2001rj,Okamoto:2002ik,Bartolo:2005fp},
\bea
 \frac35\zeta =
 \Phi_{H\rm{md}}+f_{NL}\Phi^2_{H\rm{md}}+g_{NL}\Phi^3_{H\rm{md}} + \cdots
\,.
 \eea

The primordial bispectrum and trispectrum are then given by
Eqs.~(\ref{fNLmultifielddefn}) and~(\ref{tauNLgNLdefn}), where the non-linearity
parameters $f_{NL}$ and $g_{NL}$, given in Eqs.~(\ref{fNLmultifield})
and~(\ref{gNLmultifield}), reduce to Eqs.~(\ref{fNL1field}) and~(\ref{gNL1field})
respectively, and $\tau_{NL}$ given in Eq.~(\ref{tauNLmultifield}) reduces to
\bea
%\label{fNL1field} f_{NL}&=&\frac56\frac{N''}{(N')^2}\,, \\
\label{tauNL1field}
 \tau_{NL}&=&\frac{(N'')^2}{(N')^4}=\frac{36}{25}f_{NL}^2\,.
% \\
%\label{gNL1field} g_{NL}&=&\frac{25}{54}\frac{N'''}{(N')^3}\,,
\eea
Notice that the bispectrum depends linearly on $\zeta_2$ while the trispectrum has both a
quadratic dependence upon $\zeta_2$ and a linear dependence on $\zeta_3$. Thus
$\tau_{NL}$ is proportional to $f_{NL}^2$ (shown in \cite{Okamoto:2002ik} using the
Bardeen potential, and in \cite{Boubekeur:2005fj} using this notation). However the
trispectrum could be large even when the bispectrum is small because of the $g_{NL}$ term
\cite{Okamoto:2002ik,Sasaki:2006kq}.

\subsubsection{Inflaton scenario}

In the case of standard single field inflation, where the primordial
curvature perturbation is generated solely by the inflaton field, we
can calculate the non-linearity parameters $f_{NL}$ and $g_{NL}$ in
terms of the slow-roll parameters at Hubble-exit. Because the
large-scale perturbations are adiabatic, $\zeta$ is non-linearly
conserved on large scales
\cite{Lyth:2003im,Lyth:2004gb,Langlois:2005ii} and the derivatives of
the expansion, $N'$, $N''$ and $N'''$ can be calculated at
Hubble-exit. Using the definition (\ref{localN}), we find
\bea N'&=& \frac{\bar{H}}{\dot{\bar{\varphi}}}
\simeq \frac{1}{\sqrt{2}}\frac{1}{\Mp}\frac{1}{\sqrt{\epsilon}}\sim\mathcal{O}\left(\epsilon^{-\frac12}\right)\,,  \\
N''&\simeq&-\frac12\frac{1}{\Mp^2}\frac{1}{\epsilon}(\eta-2\epsilon)\sim\mathcal{O}\left(1\right)\,,  \\
N'''&\simeq&\frac{1}{\sqrt{2}}\frac{1}{\Mp^3}\frac{1}{\epsilon\sqrt{\epsilon}}\left(\epsilon\eta-\eta^2+\frac12\xi^2\right)
\sim\mathcal{O}(\epsilon^{\frac12})\,, \eea
where we have used the potential slow roll parameters
\bea \epsilon&\equiv&\frac{\Mp^2}{2}\left(\frac{V'}{V}\right)^2 \,, \\
\eta&\equiv&\Mp^2\frac{V''}{V}\,, \\ \xi^2&\equiv& \Mp^4 \frac{V'V'''}{V^2}\,. \eea
Hence the non-linearity parameters for single field inflation
(\ref{fNL1field}--\ref{gNL1field}) are given by
\bea f_{NL}&=&\frac56(\eta-2\epsilon)\,, \\ \tau_{NL}&=&(\eta-2\epsilon)^2\,, \\
g_{NL}&=&\frac{25}{54}\left(2\epsilon\eta-2\eta^2+\xi^2\right)\,. \eea

Note however that we have not calculated the full bispectrum and trispectrum at leading
order in slow roll, because we assumed that the initial field fluctuations were Gaussian.
If we included the contribution from the non-Gaussianity of the fields at Hubble exit,
the bispectrum would have one extra term (\ref{deftrispectrum}) and the trispectrum would
have two extra terms, (\ref{trispectrumfull}). The extra term for the trispectrum is at
the same order in slow roll, because \cite{Seery:2005gb} $
B^{ABC}(\bkone,\bktwo,\bkthree)\sim \mathcal{O}(\epsilon^{\frac12}) $
and the second term of (\ref{trispectrumfull}) is also of the same order. However Seery,
Lidsey and Sloth find \cite{Seery:2006vu},
$T^{ABCD}(\bkone,\bktwo,\bkthree,\bkfour)\sim\mathcal{O}(1) $ which means the first term
of (\ref{trispectrumfull}) is suppressed by one less order in slow roll then the other
three terms. However they find the contribution of this term is still too small to be
observable, even in the multiple field case \cite{Seery:2006vu}. All of these extra terms
from the non-Gaussian field fluctuations are momentum dependent, while all of the
non-linearity parameters are independent of momentum.

In single field inflation, $\zeta$ is conserved at all orders on superhorizon scales.
Therefore no evolution of the bispectrum and trispectrum is possible after Hubble exit.
Hence neither will be detectable in cosmic microwave background or large scale structure
experiments, but from the 21 cm background there is the possibility that $f_{NL}\sim
0.01$ might be observable \cite{Cooray:2006km}.

\subsubsection{Curvaton scenario}

In the curvaton scenario a weakly-coupled field (the curvaton field, $\chi$) which is
light, but subdominant during inflation comes to contribute a significant fraction of the
energy density of the universe sometime after inflation. After it eventually decays, it
is the fluctuations in this field that produce the primordial curvature perturbation,
$\zeta$ \cite{curvaton}.

In general, the energy density of the curvaton is some function of the field value at
Hubble-exit, $\rho_\chi\propto g^2(\chi_*)$, and hence the primordial curvature
perturbation when the curvaton decays is of local form (\ref{fandgdefn}).
%In the simplest scenario, consistent with a weakly-coupled field, $g$
%is a linear function.
%
In the sudden-decay approximation the non-linearity parameters are
\cite{Bartolo:2004ty,Lyth:2005fi}
\bea f_{NL} &=& \frac{5}{4r}\left(1+\frac{gg''}{g'^2}\right)-\frac53-\frac{5r}{6}\,, \eea
and \cite{Sasaki:2006kq}
\bea
g_{NL}&=&\frac{25}{54}\left[\frac{9}{4r^2}\left(\frac{g^2g'''}{g'^3}+3\frac{gg''}{g'^2}\right)
-\frac9r\left(1+\frac{gg''}{g'^2}\right)+\frac12\left(1-9\frac{gg''}{g'^2}\right)+10r+3r^2\right]\,,
\eea
and $\tau_{NL}$ satisfies (\ref{tauNL1field}),
where the parameter $r$, is given by
\bea
 r = \left[
 \frac{3{\rho}_{\chi}}{3{\rho}_{\chi}+4{\rho}_r}
 \right]_{\rm decay}\,,
\eea
where ${\rho}_{\chi}$ is the density of the curvaton field and ${\rho}_r$ is the density
of radiation and hence $r$ satisfies $0< r\leq 1$, .

One can obtain significant non-Gaussianity if the curvaton does not dominate the total
energy density of the Universe when it decays, $r\ll 1$, in which case we have \bea
f_{NL}&\simeq&\frac{5}{4r}\left(1+\frac{gg''}{g'^2}\right)
\,, \\
g_{NL} &\simeq& \frac{25}{24r^2}\left(\frac{g^2g'''}{g'^3}+3\frac{gg''}{g'^2}\right)\,.
\eea
One obtains a significant bispectrum, $f_{NL}\gg1$ for $r\ll 1$ if
$gg''/g^{\prime2}\neq-1$. On the other hand if
$gg''/g^{\prime2}\simeq-1$ the bispectrum can be small even for
$r\ll1$ \cite{Enqvist:2005pg} and the first signal of non-Gaussianity
could come from the trispectrum through $g_{NL}\gg1$
\cite{Sasaki:2006kq}.

\section{Conclusions}

In this paper we have given a general expression characterising the distribution of the
primordial curvature perturbation due to an initial distribution of scalar field
fluctuations during multiple-field inflation, using the $\delta N$-formalism.

We have given expressions for the power spectrum, bispectrum and trispectrum including
all terms at leading order in a perturbative expansion, allowing for scale dependence and
cross-correlations of the fields. In particular the connected part of the primordial
trispectrum consists of four terms each with a different momentum dependence. One term
depends upon the intrinsic connected part of the trispectrum of the field fluctuations
and one term depends upon the bispectrum of the fields.

At lowest order in a slow-roll expansion the field fluctuations shortly after Hubble-exit
are expected to be Gaussian and scale-invariant.
In this case Lyth and Rodriguez \cite{Lyth:2005fi} showed that the bispectrum from
multiple-field inflation can be parameterised by a single non-linearity parameter
$f_{NL}$ dependent upon the second-derivatives of the local expansion, $N$, with respect
to the field values, given in Eq.~(\ref{fNLmultifield}).
The connected part of the trispectrum can be parameterised by two
further non-linearity parameters, $\tau_{NL}$ and $g_{NL}$, given in
Eqs.~(\ref{tauNLmultifield}) and~(\ref{gNLmultifield}). $\tau_{NL}$ is
another function of the second derivatives of $N$
\cite{Alabidi:2005qi}, whereas $g_{NL}$ is a dependent on the third
derivative. In the particular case where only one field generates the
primordial perturbation we have $\tau_{NL}\propto f_{NL}^2$
\cite{Okamoto:2002ik,Boubekeur:2005fj}.  However $g_{NL}$ can give
rise to a connected part of the trispectrum at the same order
\cite{Sasaki:2006kq} and hence constraints on the primordial
bispectrum do not necessarily constrain the primordial trispectrum in
such models.

In the case of standard single field inflation the trispectrum will be too small to ever
be observed, but in alternative models such as DBI inflation \cite{Huang:2006eh} or the
curvaton scenario \cite{Sasaki:2006kq} the trispectrum may be observable and could be an
important test of such models.

\textit{Note added:} While writing up this work, similar results appeared on the arXiv
\cite{Seery:2006js}.

\section*{Acknowledgements}

The authors are grateful to Jussi Valiviita, Kishore Ananda and Antony Lewis for
comments. We thank the organizers of the Benasque workshop, \textit{Modern cosmology,
Inflation, CMB and LSS}, August 2006, where this work was initiated. CB acknowledges
financial support from the EPSRC. MS is supported in part by JSPS Grant-in-Aid for
Scientific Research (S) No.~14102004, (B) No.~17340075, and (A) No.~18204024.

%MS is supported in part by JSPS Grant-in-Aid for Scientific Research (S) No.~14102004 and
%(B) No.~17340075.

\end{document}